# PC-bzip2: a phase-space continuity enhanced lossless compression algorithm for light field microscopy data

Changqing Su[1,†], Zihan Lin[2,†], You Zhou[3], Shuai Wang[4,5], Yuhan Gao[4,5], Chenggang Yan[4], Bo Xiong[1,*]

[1]National Engineering Laboratory for Video Technology (NELVT), Peking University, Beijing, 100871, China

[2]School of Automation, Hangzhou Dianzi University, Hangzhou 310018, China

[3]School of Electronic Science and Engineering, Nanjing University, Nanjing 210023, China

[4]School of Communication Engineering, Hangzhou Dianzi University, Hangzhou 310018, China

[5]Lishui Institute of Hangzhou Dianzi University, Hangzhou 323000, China

*To whom correspondence should be addressed.

† Contributed equally.

## Abstract

**Motivation:** Light-field fluorescence microscopy (LFM) is a powerful elegant compact method for long-term high-speed imaging of complex biological systems, such as neuron activities and rapid movements of organelles. LFM experiments typically generate terabytes image data and require a huge number of storage space. Some lossy compression algorithms have been proposed recently with good compression performance. However, since the specimen usually only tolerates low power density illumination for long-term imaging with low phototoxicity, the image signal-to-noise ratio (SNR) is relatively low, which will cause the loss of some efficient position or intensity information by using such lossy compression algorithms. Here, we propose a phase-space continuity enhanced bzip2 (PC-bzip2) lossless compression method for LFM data as a high efficiency and open-source tool, which combines GPU-based fast entropy judgement and multicore-CPU-based high-speed lossless compression.

**Results:** Our proposed method achieves almost 10% compression ratio improvement while keeping the capability of high-speed compression, compared with original bzip2. We evaluated our method on fluorescence beads data and fluorescence staining cells data with different SNRs. Moreover, by introducing the temporal continuity, our method shows the superior compression ratio on time series data of zebrafish blood vessels.

**Availability and implementation:** The codes, the documentation, and example data are available on an open source at: https://github.com/Onetism/LightFieldMicroscopy_PC-bzip2

**Contact:** boxiong11@outlook.com

**Supplementary information:** Supplementary data are available at Bioinformatics online.

## 1 Introduction

Light field fluorescence microscopy (LFM) serves as an elegant compact solution to long-term high-speed volumetric microscopy due to its low-photobleaching and simultaneous 3D imaging, and is suitable for biological applications such as neuron activity observing (Prevedel, R. et al., 2014; Skocek, O. et al., 2018; Zhang, Z. et al., 2021) and high-speed organelle tracking (Wu, J. et al., 2021; Xiong, B. et al., 2021; Wagner, N. et al., 2019). The data acquired from LFM experiments are with multiple dimensions, including 4D phase space and one temporal dimension. By this, the generated multi-dimensional data typically reach terabytes, with the increasing of experimental size (Amat, F. et al., 2015). Such a data production speed brings huge challenges to data storage (Li, A. et al., 2019; Andreev, A. et al., 2020). In this case, a suitable compression algorithm will greatly ease the storage pressure. Lossy compression algorithms (Shukla, K K. et al., 2011; Rabbani, M. et al., 1991; Zheng, S. et al., 2021) often provide extremely high compression ratios but will introduce uncontrolled information loss (Cromey, D. W., 2013). Lossless compression which exactly preserves the original information may have its superiority in these situations. Biomedical imaging data always require lossless compression to avoid legal issues and wrong diagnosis (Bairagi, V. K., 2015; Kaur, H. et al., 2015; Matos, L. M. et al., 2015; Nagoor O H. et al., 2021). Generally used lossless image compression methods including PNG, FLIF (Sneyers J. et al., 2016), bzip2 and KLB (Amat, F. et al., 2015) cannot take full advantage of the redundancy of LFM data for optimization. In contrast, B3D (Balázs, B. et al., 2017) uses 3D spatial continuity to increase the compression ratio, but the LFM data has redundancy in 4D phase space (Wu, J. et al., 2021; Waller, L. et al., 2012). Deep generative models based lossless compression methods such as VAEs (Townsend J. et al., 2019; Townsend J. et al., 2020) and flow-based generative models (Kingma F. et al., 2020; Ho, J. et al., 2019; Hoogeboom, E. et al., 2019) need a large amount of training data to tune the model parameters and always limit the input data size in order to control the model size. Thereby a large-scale image (e.g., 2048*2048 pixels) needs to be split into multiple pieces for compression, resulting in over 10 minutes compression time currently. The performance of VAE based method is restricted by the lower bound, which means the compression on a single data example may shows bad performance (Hoogeboom, E. et al., 2019). Flow-based lossless compression methods, based on fully observed models, is suitable for a small number of communicating samples (Townsend J. et al., 2020). Generative Adversarial Networks (Goodfellow, I. et al., 2014) based compression methods could obtain high fidelity compression (Mentzer, F. et al., 2020; Wu L. et al., 2020; Kudo S. et al., 2019) but it could not be used for lossless compression, since it doesn't optimize for likelihood (Hoogeboom, E. et al., 2019).

In light field photography for macro scenes, a light field image could be compressed according to its four-dimensional structure by making full use of the angular continuity, which refers to the utilization of time continuity in video compression techniques (Magnor, M. et al., 2000; Wu, G. et al., 2017). Besides, coding a part of sub-aperture images of light field data with depth map could also greatly improve the efficiency of light field compression (Huang, X. et al., 2019). But both methods are achieved on 8bit RGB light field images with lossy compression, since the loss of some information (e.g., tiny structures) in macro photographic data is tolerable. Therefore, existing methods cannot be directly applied to the lossless compression of microscopy data with low signal-to-noise ratio (SNR) and high dynamic range (HDR) such as fluorescence microscopy data.

Here we propose a 4D phase-space continuity enhanced bzip2 (PC-bzip2) lossless compression method, as a light-weight and high-throughput compression tool for LFM data. By applying a prediction based on 4D phase-space continuity and GPU based fast entropy judgement, we could get a predicted image with smaller size and then compress the image with multicore-CPU-based high-speed lossless compression method. Compared with original bzip2, we achieve almost 10% compression ratio improvement while keeping the capability of high-speed compression. A fast MATLAB interface and ImageJ plug-in were provided for ease of use. To demonstrate the performance of our method, we tested on fluorescence beads data and different types of cells data under different light conditions. Moreover, we showed the superior compression ratio of our method on time series data of zebrafish blood vessels by introducing temporal continuity prediction.

## 2 Methods

### 2.1 PC-bzip2 framework for high-speed lossless LFM data compression

We built a light field fluorescence microscope based on widefield microscope, by inserting a microlens array to the image plane of tube lens (**Fig. 1a**). In this way, the light field of the sample will be formed behind the microlens array which then will be relayed to the image sensor for recording (**Fig. 1a**). A sample zebrafish image captured by the light field microscope shows the basic structure of the data. The LFM data has a four-dimensional structure with not only the spatial continuity but also angular continuity. Angular continuity characterizes the relationship between adjacent pixels in a microlens, while spatial continuity characterizes the relationship between adjacent microlens pixels at the same position. Traditional image compression algorithms mainly make predictions by using adjacent pixels. However, the predictors of light field are no longer limited to this constraint.

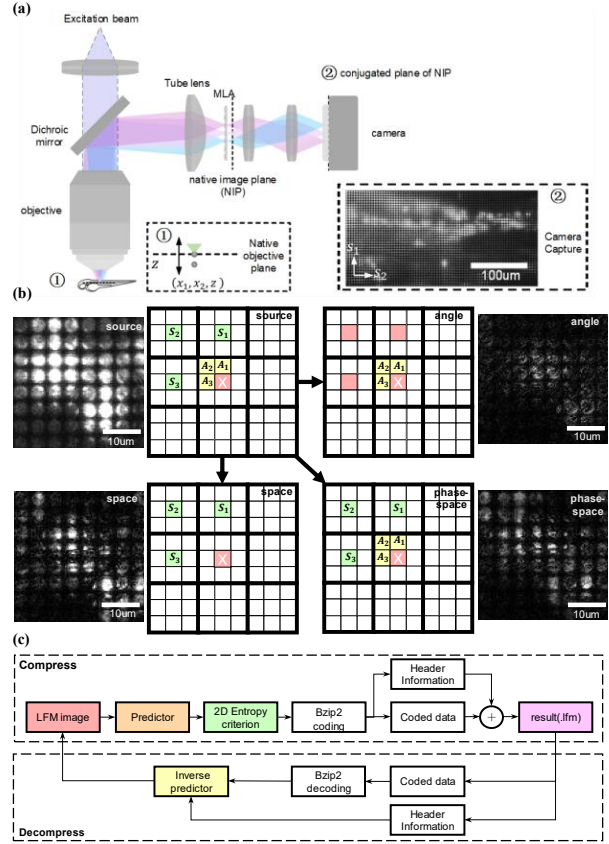

Fig. 1 PC-bzip2 framework for high-speed lossless LFM data compression. (a) Schematic of the light field microscopy system and LFM data structure. (b) Schematic of the predictors based on spatial continuity, angular continuity, and 4D phase-space continuity. (c) A complete framework including PC-bzip2 compression and decompression process. The compression pipeline consists of prediction part, 2D image entropy criterion, multi-core CPU accelerated bzip2 coding, and packing header information and coded data. The decompression pipeline consists of unpacking header information, multi-core CPU accelerated bzip2 decoding, and inverse prediction.

In this case, the angular or spatial continuity can be applied alone for prediction. What's more, we can make full use of the continuity in both angular and spatial domains, which is named phase-space continuity, to make predictions for LFM data (**Fig 1b**), thereby theoretically improving the compression ratio. Therefore, based on the phase-space continuity, we propose a pc-bzip2 compression framework for lossless LFM data compression (**Fig 1c**). The main idea of this framework is to add a prior knowledge-based predictor for the LFM data before the multi-core CPU accelerated bzip2 compression. In addition, we propose a two-dimensional image entropy criterion to determine which predictor is used to optimize the compression ratio. These two parts, as the pre-processing unit before bzip2 compression, can ensure that each compressed image has as little information redundancy as possible. Finally, the header information including the selected predictor information and the coded data can be packaged and stored out to the file. In the decompression process, after multi-core CPU accelerated bzip2 decompression, we only need to select the corresponding inverse predictor according to the header information to restore the original LFM data (**Fig. 1**).

To sum up, the proposed PC-bzip2 lossless compression pipeline can be decomposed into four parts: prediction part, 2D image entropy criterion, multi-core CPU accelerated bzip2 coding, and packing header information and coded data. We describe the details of each part in the following. The detailed schematic of algorithm is shown in **Fig. 2**.

### 2.1.1 Prediction Part

The prediction of pixel values makes full use of the continuity between pixels to reduce redundant information. As four-dimensional structure data, LFM image has 4D phase-space continuity. We separately use spatial continuity, angular continuity and 4D phase-space continuity to predict the pixel value and compare their performance. Using spatial continuity alone means the pixel value is predicted by the left, top and top left neighboring pixels. Using angular continuity alone means the pixel value is predicted by the pixels in the same position of the left, top and top left neighboring microlenses. Using 4D phase space continuity means the pixel value is not only related to neighboring pixels but also related to the pixels in neighboring microlenses, as shown in **Fig. 1b**.

### 2.1.2 Two-dimensional image entropy criterion

A two-dimensional image entropy criterion is used to determine whether the prediction of pixel value will improve compression ratio or not. After we apply the predictor on the input LFM image, we realign the 16bit predicted image as an 8bit data string and do a GPU-accelerated BWT (Burrows-Wheeler transform) on the 8bit data string. The GPU-accelerated BWT is a little bit different from the general BWT. From input to output, the general BWT includes three main steps: generating all rotations, sorting all rotations into lexical order, and taking the last column of all rotations. Different from the general BWT, when sorting all rotations, the GPU-accelerated BWT only considers the first character of each rotation. And if the first character of two rotations is the same, we keep the sorting order of these two rotations consistent with the generation. In this way, parallel acceleration can be achieved and the time to determine whether the prediction is effective can be greatly reduced. After that, the 2D image entropy of the transformed data will be calculated. Here, we suppose the transformed 8bit data string as $S_8 = [x_1, x_2, ... x_n]$ ($x_i$ is an 8bit integer). In order to retain the sequential information of adjacent characters in the data, which is an essential part of move-to-front transform in the subsequent compression, we combine two adjacent 8bit integers into a 16bit integer, i.e., the data string is converted to 16bit data string $S_{16} = [x_1x_2, x_2x_3, ... x_{n-1}x_n]$ ($x_ix_j$ is an 16bit integer). Then, the 2D image entropy of the transformed 8bit data string $E(S_8)$ could be calculated as:

$$E(S_8) = \sum_{t=0}^{65535} \frac{P_t}{n-1} log(\frac{P_t}{n-1}) \qquad (1)$$

where $t$ is the value of 16bit integer in $S_{16}$, $P_t$ represents the frequency of 16bit integer value $t$, and $n$ is the length of 8bit data string $S_8$. A detailed schematic of fast calculation of 2D image entropy is shown in **Fig 2c**.

### 2.1.3 Multi-core CPU accelerated bzip2 coding

After confirming that the prediction can improve the compression ratio through two-dimensional image entropy, the predicted image is then compressed with multicore CPU accelerated bzip2 algorithm.

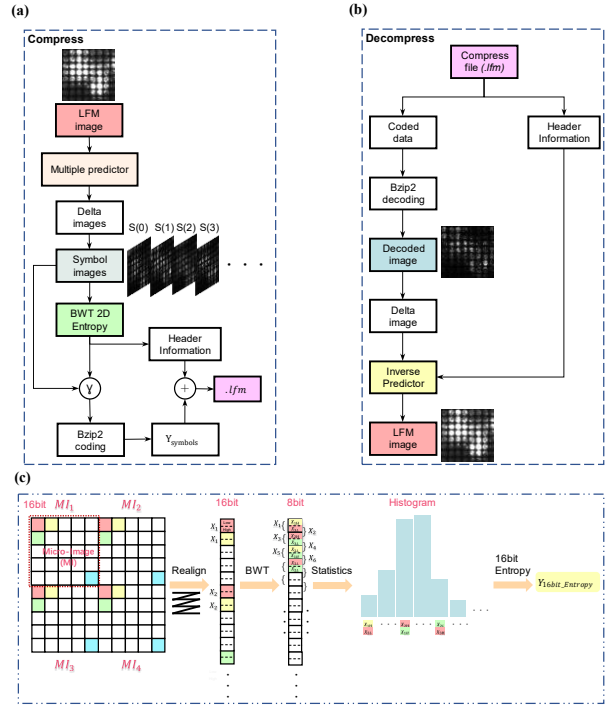

Fig. 2 The pipeline of the pc-bzip2 lossless compression algorithm. (a) The pipeline of the pc-bzip2 lossless compression with typical examples of the output for the main step. Multiple predictors step means to perform all prediction methods individually and symbol images step means to convert the delta images to full positive images. (b) The pipeline of the pc-bzip2 decompression with typical examples of the output for the main step. We need to decide which inverse prediction method to use based on the header information. The decoded image here is one of the symbol images in compression pipeline. (c) The schematic of the two-dimensional image entropy calculation. In BWT transformation, the realigned 16bit image sequence is regarded as 8bit image sequence by splitting each 16bit value into high and low 8bits. After BWT transformation, every two adjacent 8bits will be formed into 16bits for histogram statistics.

Here, we reused the KLB code for our own purpose[7]. KLB is currently one of the most advanced methods for bzip2 acceleration. The main idea of KLB is to split data into multiple blocks and then compress them in parallel, getting the utmost out of the multi-core design of advanced computers. The KLB compression process consists of three steps. First, the data to be compressed is divided into some blocks. The block number is corresponding to the number of threads supported by the computer processor, representing the maximum number of parallel operations. Second, multiple blocks are simultaneously compressed by bzip2 algorithm, which is the main reason for speed improvement of KLB compared to general bzip2. Finally, the compression result of each block in the previous step are written into the output file in turn, and the relevant information for image and the block information are packed at the same time.

### 2.1.4 Packing header information and coded data

Since we use different predictors for different images according to the 2D image entropy judgement result, we need to pack the predictor information into the final compressed file. Here, the predictor information is represented by an 8bit unsigned integer in the header information of final compressed file. The mapping table of predictor can be viewed in **Fig 3**.

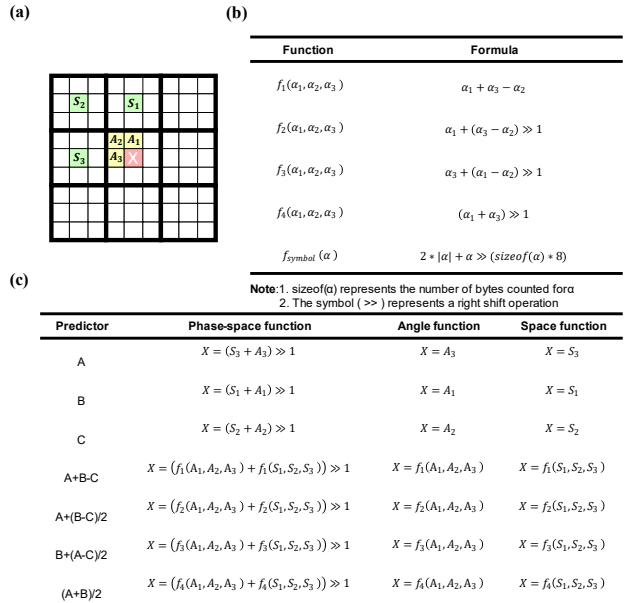

Fig 3 Main functions used in this work for LFM data compression. (a) Schematic of LFM data structure. S1, S2, S3, and X are adjacent in spatial plane, A1, A2, A3, and X are adjacent in angular plane. (b) The functions used in the predictors. The first four are common prediction functions in JPEG based on adjacent pixel information and the last one is to map all predicted values to a positive interval. (c) Lossless predictors for X, based on the schematic in (a). Phase-space function means to use 4D phase-space continuity to predict X, angle function means that only angle continuity is used to predict X and space function means using spatial continuity alone to predict X.

## 3  Results

We investigated different predictors in the compression algorithm. The schematic of the LFM data is shown in **Figure 3a**. Among the 7 types of predictors mainly applied in traditional compression methods (such as PNG and JPEG), 4 of them utilize more than one adjacent pixel for predictor calculation (**Fig. 3b**). We then extend these predictors to the LFM data (**Fig. 3c**), including angle prediction, space prediction, and phase-space prediction.

To verify the performance of our pc-bzip2 compression experimentally, we imaged the fluorescent beads and MCF-10A cells separately with the LFM under the different exposure times (**Fig. 4a**), where the image of the fluorescent beads represents a typical LFM data without continuous structural information while the image of the MCF-10A cells represents a typical LFM data with better structural information. Different exposure times will cause images to have different SNRs. Then we tested the performance of all the phase-space predictors (**Fig. 3b, 3c**) on these two types of LFM images with different SNRs. We find that as the image SNR increases (i.e., more amount of information is detected within the image), the compression ratio gradually increases, which means the increase in the number of bits required for each dimension for all the predictors (**Fig. 4b, 4c**). For the images of the fluorescent beads without continuous structural information, the prediction step didn't improve the final compression ratio (**Fig. 4c**). But for the MCF-10A cells images with better continuous structure information, the prediction step can significantly increase the final compression rate (**Fig 4b**) and the maximum improvement can reach 0.77 bits/dim. In the compression tests of both fluorescent beads images and MCF-10A cells images, the two-dimensional entropy criterion can accurately predict the optimal predictors with the optimal compression ratios (**Fig. 4b, 4c**). Besides, we further tested the predictor (B+(A-C)/2) using angle prediction, space prediction, and phase-space prediction separately (**Fig. 4d**) where the phase-space predictor showed the highest compression ratio. In addition, to further demonstrate the enhancement of the compression ratio in our pc-bzip2, we compared our pc-bzip2 with KLB on the images of MCF-10A in terms of time consumption (including compression and decompression) and compression rate. With almost no additional time consumption, the compression ratio can be increased by 10% (**Fig. 4e**). This tiny time gap can be further reduced with the upgrade of the graphics processing unit (GPU). Moreover, to visually demonstrate the negative impact of lossy compression on biomedical images, we compared the compression results of larval zebra images by using lossless and lossy methods, including the display of decompressed images and grayscale histograms (**Supplementary Fig. S1**). After lossy compression, the decompressed image obviously loses a lot of local details which is often the focus of biomedical researchers. The histogram distribution of the image after lossy compression also shows that the original continuous structural information has been smeared compared to the lossless decompressed data. Therefore, researches on lossless compression are critical to biomedical images storage, especially in the face of increasingly large data. Finally, we separately compared the decompress images of the fluorescent beads image and the MCF-10A image compressed by our pc-bzip2 with their original images (**Supplementary Fig. S2**). The results show that our method achieves truly lossless performance. Since the recording of biological dynamic processes such as neuron activity by LFM usually lasts for a long time with high frame rate, massive quantities of time series data will be generated consequently. Thereby, we have extended our method to the time dimension to further improve its practicality. More details can be found **in the Supplementary materials.**

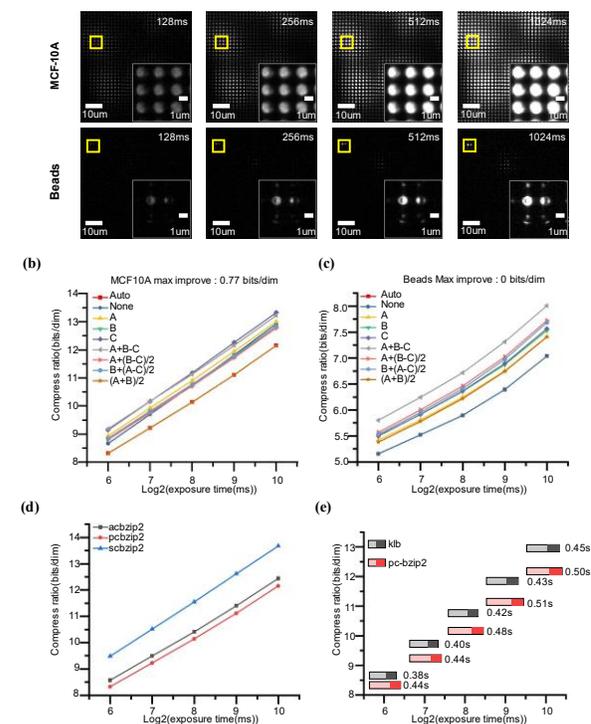

Fig. 4 Comparing different compression ratios on images of different specimens with different SNRs. (a) Two groups of samples were captured at different exposure times (the change of the exposure time resulted in different SNRs on images). The beads images were obtained by imaging green fluorescent beads with the LFM and MCF-10A images were obtained by imaging MCF-10A cells with the LFM. The image in the lower-right corner is a close-up marked by the yellow box. (b) The performance comparison on MCF-10A images with different predictors. The automatic criterion can accurately choose (A+B)/2 as the optimal predictor method and find the optimal compression ratio as well. Compared with the compression ratio without the prediction, the maximum improvement is up to 0.77bits/dim. (c) The performance comparison on beads images with different predictors. The predictor did not improve the compression ratio but the automatic criterion could always locate the best compression ratio. (d) The performance comparison of angle prediction (ac-bzip2), space prediction (sc-bzip2), and phase-space prediction (pc-bzip2) on MCF-10A images. Obviously, the phase-space prediction works best. (e) The performance comparison of KLB and pc-bzip2 in compression time, decompression time, and compress ratio respectively on MCF-10A images. The location of the rectangle corresponds to the compression ratio, its size corresponds to the sum of compression and decompression time (as indicated by the numbers next to it), the light color corresponds to the compression time and the dark color corresponds to the decompression time.

## 4 Discussion and conclusion

In summary, we have developed a 4D phase-space continuity enhanced bzip2 lossless compression method to realize high-speed and efficient LFM data compression. By adding a suitable predictor determined by two-dimensional image entropy criterion before bzip2 or KLB compression, we achieved almost 10% improvement in compression ratio with a little increase in time. We demonstrated the performance of pc-bzip2 algorithm on fluorescence beads data and cells data with different SNRs. Since the recording of multicellular organisms by LFM usually generate huge time series data, we further extended our method to the time dimension to improve its practicality. We validate the temporal extension of pc-bzip2 on time series recording of zebrafish blood vessels. Compared to the predictor in traditional compression method, we fully exploited the structure of LFM image or video to achieve high compression performance. We provided a fast MATLAB interface and ImageJ plug-in were provided for ease of use. The pc-bzip2 can become a promising and light-weight tool for any light field microscope.

Since improvement of compression ratio in pc-bzip2 is mainly determined by the redundancy of LFM data, the pc-bzip2 algorithm will not show significant improvement on all samples. Therefore, in our method, we use the two-dimensional image entropy criterion to choose a suitable predictor or directly apply bzip2 compression algorithm without predictor for the adaption to different samples. Further improvement may include extending the algorithm to spectral dimension and optimizing the time consumption of two-dimensional image entropy criterion. We believe such improvements in compression performance will bring advance data storage capacity with a light-weight tool to the broad microscopic community, facilitating the mass data storage and processing in various biomedical applications like multicellular organisms observing.

## Funding

This work has been supported by National Natural Science Foundation of China (Grant No. 62371006, 61931008, U21B2024, 62071415), National Key Research and Development Program of China under Grant (Grant No. 2020YFB1406604), Zhejiang Provincial Natural Science Foundation of China (Grant No. LDT23F01011F01, LDT23F01015F01, LDT23F01014F01), "Pioneer" and "Leading Goose" R&D Program of Zhejiang Province(2022C01068) and China Postdoctoral Science Foundation (Grant No. 2023TQ0006).